\documentclass[onecolumn,secnumarabic,amssymb, amsmath, nofootinbib,tightenlines, nobibnotes, aps, prl,epsfig]{revtex4}
\usepackage{graphicx}
\usepackage{dcolumn}
\usepackage{bm}
\begin{document}
\preprint{APS/123-QED}
\title{ Nonlinear corrections for the nuclear gluon distribution in $eA$ processes}

\author{G.R.Boroun}%
 \email{boroun@razi.ac.ir }
\author{B.Rezaei }
\altaffiliation{brezaei@razi.ac.ir}
\author{F.Abdi** }
\altaffiliation{fariba.abdi@razi.ac.ir}

\affiliation{Department of  Physics, Razi University, Kermanshah
67149, Iran}

\date{\today}
\begin{abstract}
An analytical study with respect to the nonlinear corrections for
the nuclear gluon distribution function in the next-to-leading
order approximation at small $x$ is presented. We consider the
nonlinear corrections to the nuclear gluon distribution functions
at low values of $x$ and $Q^{2}$ using the parametrization
$F_{2}(x,Q^{2})$ and using the nuclear modification factors where
they have been obtained with the
Khanpour-Soleymaninia-Atashbar-Spiesberger-Guzey model. The CT18
gluon distribution is used as baseline proton gluon density at
$Q_{0}^{2}=1.69~\mathrm{GeV}^2$. We discuss the behavior of the
gluon densities in  the next-to-leading order and the
next-to-next-to-leading order approximations at the initial scale
$Q_{0}^{2}$, as well as the modifications due to the nonlinear
corrections. We find the QCD nonlinear corrections are significant
for the next-to-leading order accuracy than the
next-to-next-to-leading order for light and heavy nuclei. The
results of the nonlinear GLR-MQ evolution equation are similar to
those obtained with the Rausch-Guzey-Klasen gluon upward and
downward evolutions
 within the uncertainties. The magnitude of the gluon distribution with the nonlinear corrections increases with a
 decrease of $x$ and an increase of the atomic number A.\\
\end{abstract}
 \pacs{***}
\maketitle
\subsection{1. Introduction}
The dynamics of parton interactions and the partonic structure of
nuclei are prime subjects of research for both particle and
nuclear physics. The formation of quark-gluon plasma inside nuclei
is explored during the very first fractions of  $\mathrm{fm}/c$ in
high-energy nuclear collisions. This probe is due to a large
momentum (or mass) $Q{\gg}\Lambda_{QCD}$ scale, which is the main
motivation for studying nuclear parton distributions. According to
the knowledge of the parton distribution functions (PDFs) of free
nucleons which comes from the measurements of deeply inelastic
scattering (DIS) in lepton-nucleon ($lN$) collisions, the program
of extracting nuclear PDFs (nPDFs) also relies on the DIS data
[1-3]. The HERA data for the free proton reached $x{\sim}10^{-5}$
in perturbative values of $Q^2$, while the DIS-measurements for
nuclear targets are bound to severely higher momentum fractions,
$x{\gtrsim}10^{-2}$.\\
In Ref.[4], the authors studied the prospects for constraining the
nuclear parton distribution functions by small-$x$ deep inelastic
scattering at the Large Hadron Electron Collider (LHeC) [5] where
its extension of the kinematic covers 4 orders of magnitude in
DIS. The effect of high-precision DIS-measurements at the LHeC in
Ref.[4] is illustrated by the ratio of the reduced, inclusive DIS
cross-sections,
$\sigma^{A}_{\mathrm{reduced}}(x,Q^2)/\sigma^{p}_{\mathrm{reduced}}(x,Q^2)$,
where
\begin{eqnarray}
\sigma_{\mathrm{reduced}}(x,Q^2)=F_{2}(x,Q^2)\Big{[}1-\frac{y^2}{1+(1-y)^2}\frac{F_{L}}{F_{2}}\Big{]},
\end{eqnarray}
where $x$, $y$ and $Q^2$ are the standard DIS variable, and $A$ is
the number of nucleons in a nuclear target. The LHeC promises the
equivalent of $1~\mathrm{fb}^{-1}$ of luminosity for
$\mathrm{ePb}$ collisions at LH(e)C energies. With its large $Q^2$
and $1/x$ range
nuclear shadowing can be measured very precisely.\\
At high energies, nuclear shadowing is controlled by coherence
effects. Namely, shadowing is possible only if the coherence time
exceeds the mean inter-nucleon spacing in nuclei and shadowing
saturates if the coherence time substantially exceeds the nuclear
radius [6-8]. Nuclear shadowing at small $x$ (i.e.,
$x{\lesssim}0.1$) is experimentally well studied by NMC [9].
Experiments at CERN and Fermilab focus especially on the region of
small values of the Bjorken variable $x$ and show a systematic
reduction of the nuclear structure function $F^{A}_{2}(x,Q^{2})/A$
with respect to the free proton structure function
$F^{p}_{2}(x,Q^{2})$. This phenomenon is known as nuclear
shadowing effect and is associated to the modification of the
target parton distributions so that
$xf_{i}^{A}(x,Q^{2})<Axf_{i}^{p}(x,Q^{2})$, $f_{i}=q,g,..$ [10].
The relation of the bound-proton PDFs with respect to free-proton
PDFs $f_{i}^{p}$ is often expressed in terms of the nuclear
modification factors
$R_{i}^{A}(x,Q^2)=f_{i}^{p/A}(x,Q^2)/f_{i}^{p}(x,Q^2)$. For a
nucleus $A$ with $Z$ protons and $N=A-Z$ neutrons, an average PDF
is obtained as
\begin{eqnarray}
f_{i}^{A}(x,Q^2)=\frac{Z}{A}f_{i}^{p/A}(x,Q^2)+\frac{N}{A}f_{i}^{n/A}(x,Q^2),
\end{eqnarray}
where $f_{i}^{p/A}$ are the PDFs of a bound proton and the neutron
contents $f_{i}^{n/A}$ are obtained from $f_{i}^{p/A}$ via isospin
symmetry [11-14]. As revealed by DIS experiments, the bound
nucleon PDFs are not the same as those of a free proton, but are
modified in a nontrivial way and obey the
Dokshitzer-Gribov-Lipatov-Altarelli-Parisi (DGLAP) evolution
[15-18], which describes how the PDFs depend on the factorization
scale
\begin{eqnarray}
Q^2\frac{{\partial}f_{i}}{{\partial}Q^2}=\sum_{j}P_{ij}{\otimes}f_{j},
\end{eqnarray}
with splitting functions $P_{ij}$ governing the scale evolution.
For evolution of the PDFs due to the evolution equation (i.e.,
Eq.(3)), a non-perturbative input at some initial scale is
required to obtain a PDF set. The baseline parton distributions of
a proton are parametrized in the following formal form
\begin{eqnarray}
f_{i}(x,Q^{2}_{0})=\alpha_{0}x^{\alpha_{1}-1}(1-x)^{\alpha_{2}}P_{i}(y,\alpha_{3},\alpha_{4},...),
\end{eqnarray}
where the coefficients $\alpha_{1}$ and $\alpha_{2}$ control the
asymptotic behavior of $f_{i}(x,Q^{2}_{0})$ in the limits
$x{\rightarrow}0$ and $1$, and $P_{i}$ is a sum of Bernstein
polynomials dependent on $y=f(x)$ which is very flexible across
the whole interval $0<x<1$. For PDFs of nuclei, an additional
dependence on the atomic mass A is required  [12,14,19]. In
Ref.[20], the authors are discussed the nuclear cross section in
terms of nuclear volume and surface contributions
\begin{eqnarray}
\sigma_{A}=A\sigma_{V}+A^{2/3}\sigma_{S}.
\end{eqnarray}
Therefore, the cross section per nucleon is assumed to be
proportional to $1/A^{1/3}$ as
\begin{eqnarray}
\frac{\sigma_{A}}{A}=\sigma_{V}+\frac{1}{A^{1/3}}\sigma_{S}.
\end{eqnarray}
If $\sigma_{V}$ and $\sigma_{S}$ depend weakly on A, the
$1/A^{1/3}$ dependence makes sense as the leading approximation.\\
A much harder task has been to determine the gluon distribution of
nucleons bound in a nucleus, i.e., the nuclear gluon distribution
($xg^{A}(x,Q^{2})$). The kinematic extension of the the electron -
Ion collider (EIC) [21,22] will allow us to examine the non-linear
dynamics at low $x$. When the gluon density becomes sufficiently
large at small $x$, one needs to take into account the effects of
gluon recombination (gluon-gluon fusion) leading to nonlinear
corrections to the DGLAP evolution equations [23-25].  Indeed the
gluon-gluon recombination processes cause that the growth of the
gluon density is slowed down at smaller values of $x$ and $Q^{2}$
(but still $Q^{2}{\gg}\Lambda^{2}_{QCD}$). In the
Gribov-Levin-Ryskin-Mueller-Qiu (GLR-MQ) approach [23,24], the
gluon recombination is addressed by analyzing so-called $"$fan$"$
diagrams, where two gluon ladders merge into a gluon or a
quark-antiquark pair. Adding these contributions to the DGLAP
equations yields the nonlinear GLR-MQ evolution equations [23,24],
where the nonlinear term tames the growth of the PDFs at small $x$
and leads to their suppression.  One of the important outcomes of
studied in Ref.[26] is the existence of the saturation scale
$Q_{s}(x)$ ($Q_{s}^{2}=Q_{0}^{2}(x/x_{0})^{-\lambda}$ where
$Q_{0}$ and $x_{0}$ are free parameters) which is a characteristic
scale at which the parton recombination effects become important.
The solution to the non-linear equation has the property of the
geometric scaling in the regime where $k<Q_{s}(x )$ whereas in the
case when $k>Q_{s}(x)$ the solution enters the linear regime,
where $k$ is
the gluon transverse momenta.\\
Effects of small-$x$ nonlinear corrections to the DGLAP evolution
equations due to gluon recombination have been extensively studied
in the literature [27-32]. Recently in Ref.[33], the authors have
considered the nonlinear GLR-MQ evolution equations for nPDFs
using the $"$brute force$"$ method in the momentum space. The
authors [33] confirmed the importance of the nonlinear corrections
for small $x{\lesssim}10^{-3}$, whose magnitude increases with a
decrease of $x$ and an increase of the atomic number A.  This
paper is organized as follows. In the next section the
 theoretical formalism is presented, including the GLR-MQ evolution equation.
  In section 3, we present
  a detailed analytical analysis and our main results for the nuclear gluon density and
  predictions of the non-linear effects at higher order accuracy. In the last
section we summarize our findings.\\

\subsection{2. Formalism}

The nonlinear corrections in the GLR-MQ evolution equations for
nPDFs are defined by the following forms\footnote{For future
discussion please see the Appendix.}
\begin{eqnarray}
\frac{\partial{xg^{A}(x,Q^{2})}}{\partial{\ln}Q^{2}}=\frac{\partial{xg^{A}(x,Q^{2})}}{\partial{\ln}Q^{2}}|_{\mathrm{DGLAP}}
-\frac{81}{16}\frac{\alpha_{s}^{2}(Q^{2})}{\mathcal{R}^{2,A}Q^{2}}\int_{\chi}^{1}\frac{dz}{z}[\frac{x}{z}g^{A}(\frac{x}{z},Q^{2})]^{2}
\end{eqnarray}
 and
\begin{eqnarray}
\frac{\partial{xq_{s}^{A}(x,Q^{2})}}{\partial{\ln}Q^{2}}=\frac{\partial{xq_{s}^{A}(x,Q^{2})}}{\partial{\ln}Q^{2}}|_{\mathrm{DGLAP}}
-\frac{27\alpha_{s}^{2}(Q^{2})}{160\mathcal{R}^{2,A}Q^{2}}[xg^{A}(x,Q^{2})]^{2},
\end{eqnarray}
where
$\frac{\partial{f_{i}(x,Q^{2})}}{\partial{\ln}Q^{2}}|_{\mathrm{DGLAP}}$
for the parton distributions refer to the standard DGLAP evolution
equations. Here $\mathcal{R}$ is the characteristic radius of the
gluon distribution in the hadronic target. $\mathcal{R}^{A}$ for a
nuclear target with the mass number A is defined by
$\mathcal{R}^{A}=2~\mathrm{GeV^{-1}}{\times}A^{1/3}$ [33]. The
value of $2~\mathrm{GeV^{-1}}$ depends on how the gluons have a
hotspot-like structure within the nucleon. Here
$\chi=\frac{x}{x_{0}}$ and $x_{0}$ is the boundary condition that
the gluon distribution joints smoothly onto the linear region. The
second terms in the right-hand sides of Eqs.(7) and (8) are
expected to become important and related to the recombination of
the gluons in the low-$x$ region, when the gluon density is very
large. This is known as the phenomenon of gluon saturation.\\
Since the parton distributions in bound and free protons are
different, $f^{A}(x,Q^2){\neq}f(x,Q^2)$, therefore the ratio of
structure functions is observed to deviate clearly from unity. The
nuclear modifications at $x{\lesssim}0.1$ are referred to as
shadowing. The nuclear structure function $F_{2}^{A}$, in the
QCD-improved parton model (in leading order (LO) of $\alpha_{s}$,
or in the DIS-scheme in any higher order), can be written in terms
of its parton distributions as
\begin{eqnarray}
F_{2}^{A}(x,Q^2)=\sum_{i=u,d,s,...}e_{q}^{2}\Big{[}xq_{i}^{A}(x,Q^2)+x\overline{q}_{i}^{A}(x,Q^2)\Big{]},
\end{eqnarray}
where $e_{q}$ is the quark charge, and $q^{A}(\overline{q}^{A})$
is the quark (antiquark) density in the nucleus A. The nuclear
structure function, with assumed flavor symmetric antiquark
distributions, becomes a summation of valence quark and antiquark
distributions
\begin{eqnarray}
F_{2}^{A}(x,Q^2)=\frac{x}{9}\Big{[}4u_{v}^{A}(x,Q^2)+d_{v}^{A}(x,Q^2)
+12\overline{q}^{A}(x,Q^2)\Big{]}.\nonumber\\
\end{eqnarray}
The nonlinear equations (i.e., Eqs.(7) and (8)) show that the
strong rise that is corresponding to the linear QCD evolution
equation at small-$x$ and $Q^2$ can be tamed by screening effects.
After successive integrating of both sides of Eqs.(7) and (8) with
respect to $\ln{Q^2}$ and some rearranging, we find the nonlinear
 distribution functions in terms of the linear by the following
 forms
\begin{eqnarray}
\int_{Q_{0}^{2}}^{Q^2}d[xg^{A}(x,Q^{2})]=\int_{Q_{0}^{2}}^{Q^2}[dxg^{A}(x,Q^{2})]|_{\mathrm{DGLAP}}
-\int_{Q_{0}^{2}}^{Q^2}\frac{81}{16}\frac{\alpha_{s}^{2}(Q^{2})}{\mathcal{R}^{2,A}Q^{2}}d{\ln}Q^2\int_{\chi}^{1}\frac{dz}{z}[\frac{x}{z}g^{A}(\frac{x}{z},Q^{2})]^{2}
\end{eqnarray}
 and
\begin{eqnarray}
\int_{Q_{0}^{2}}^{Q^2}d[xq_{s}^{A}(x,Q^{2})]=\int_{Q_{0}^{2}}^{Q^2}d[xq_{s}^{A}(x,Q^{2})]|_{\mathrm{DGLAP}}
-\int_{Q_{0}^{2}}^{Q^2}\frac{27\alpha_{s}^{2}(Q^{2})}{160\mathcal{R}^{2,A}Q^{2}}[xg^{A}(x,Q^{2})]^{2}d{\ln}Q^2.
\end{eqnarray}
Integrating the first terms in the left and right hands of
Eqs.(11) and (12) and using the linear and nonlinear initial
conditions $xf^{A}_{i}(x,Q_{0}^{2})$ (given by Eqs.(15), (19) and
(20) below), we find the nonlinear corrections (NLCs) to the
parton distribution functions by the following forms
\begin{eqnarray}
xg^{A,NLC}(x,Q^{2})=xg^{A,NLC}(x,Q_{0}^{2})+[xg^{A}(x,Q^{2})-xg^{A}(x,Q_{0}^{2})]-\int_{Q_{0}^{2}}^{Q^2}\frac{81}{16}\frac{\alpha_{s}^{2}(Q^{2})}{\mathcal{R}^{2,A}Q^{2}}d{\ln}Q^2\int_{\chi}^{1}\frac{dz}{z}[\frac{x}{z}g^{A}(\frac{x}{z},Q^{2})]^{2},
\end{eqnarray}
 and
\begin{eqnarray}
xq_{s}^{A,NLC}(x,Q^{2})=xq_{s}^{A,NLC}(x,Q_{0}^{2})+[xq_{s}^{A}(x,Q^{2})-xq_{s}^{A}(x,Q_{0}^{2})]-\int_{Q_{0}^{2}}^{Q^2}-\frac{27\alpha_{s}^{2}(Q^{2})}{160\mathcal{R}^{2,A}Q^{2}}[xg^{A}(x,Q^{2})]^{2}d{\ln}Q^2.
\end{eqnarray}
Here $xf_{i}^{A}(x,Q^2)$ and $xf_{i}^{A}(x,Q_{0}^2)$ are the
linear parton distribution functions at the scales of $Q^2$ and
$Q_{0}^{2}$ respectively, and obtained from the coupled DGLAP
evolution equations using the modified nuclear distribution
functions at the initial scale\footnote{The linear gluon
distributions  at the higher order approximations are discussed
using the Laplace transform at $Q^2$ scale in Sec.3. }. The
initial nuclear parton distributions are provided at a fixed
$Q^{2}$ (${\equiv}Q_{0}^{2}$), due to a free nucleon distribution
function, $f_{i}(x,Q_{0}^{2})$, and a multiplicative nuclear
modification factor, $w_{i}(x,A,Z)$, as
\begin{eqnarray}
f^{A}_{i}(x,Q_{0}^{2})=w_{i}(x,A,Z)f_{i}(x,Q_{0}^{2}).
\end{eqnarray}
The nuclear modification is based on the QCD analysis available in
the literature [14, 19, 34-38], and assume the following
modification function
\begin{eqnarray}
w_{i}(x,A,Z)=1+\Big{(}1-\frac{1}{A^{\alpha}}\Big{)}\frac{a_{i}(A,Z)+H_{i}(x)}{(1-x)^{\beta_{i}}},
\end{eqnarray}
where $H_{i}(x)=b_{i}(A)x+c_{i}(A)x^2 +d_{i}(A)x^3$ is in the
cubic type. An advantage of the cubic form with the additional
term $d_{i}$ in contrast to a quadratic-type function, i.e.,
without $d_{i}$, is that the weight function becomes flexible
enough to accommodate both shadowing and anti-shadowing in the
valence quark
distributions [14].\\
\begin{figure}
\centerline{
\includegraphics[width=0.58\textwidth]{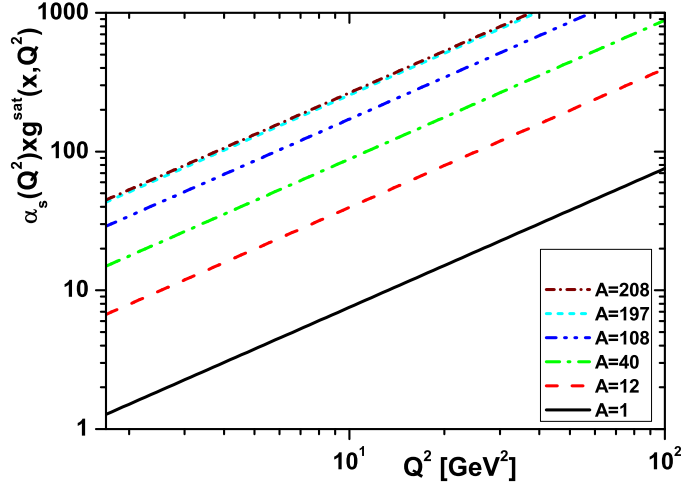}}
\caption{ Results of $\alpha_{s}xg^{A}_{\mathrm{sat}}$ for
different values of $Q^2$ in a wide range of nuclei including
C-12, Ca-40, Ag-108, Au-197, Pb-208 and the free
proton.}\label{Fig1}
\end{figure}
The nonlinear corrections enter both the gluon and the sea-quark
distributions at small $x$ through (i) modifications of the
initial distributions and (ii) the presence of additional
nonlinear terms in the $Q^2$-evolution equations. To study the
possible importance of nonlinear corrections, we base our initial
gluon and singlet distribution $xg^{NLC}(x,Q_{0}^{2})$ and
$xq_{s}^{NLC}(x,Q_{0}^{2})$ by imposing nonlinear corrections on
linear distribution functions. The nonlinear corrections to the
gluon distribution, at the initial scale $Q_{0}^2$, is obtained
from the results in Ref.[39] as\footnote{For future discussion
please see Ref.[39].}
\begin{eqnarray}
xg^{A,\mathrm{NLC}}(x,Q_{0}^2)=xg^{A}(x,Q_{0}^2)\Big{\{}1+\theta(x_{0}-x)\Big{[}
xg^{A}(x,Q_{0}^2)-xg^{A}(x_{0},Q_{0}^2)
\Big{]}/xg^{A}_{\mathrm{sat}}(x,Q_{0}^{2})\Big{\}}^{-1},
\end{eqnarray}
where
\begin{eqnarray}
xg^{A}_{\mathrm{sat}}(x,Q^{2})=\frac{16\mathcal{R}^{2,A}Q^2}{27{\pi}\alpha_{s}(Q^2)}.
\end{eqnarray}
The nonlinear terms in the right-hand side of evolution equations
(i.e., Eqs.(7) and (8)) are defined by $xg^{A}_{\mathrm{sat}}$,
and this is the value of the gluon which would saturate the
unitarity limit in the leading shadowing approximation.
\begin{figure}
\centerline{
\includegraphics[width=0.58\textwidth]{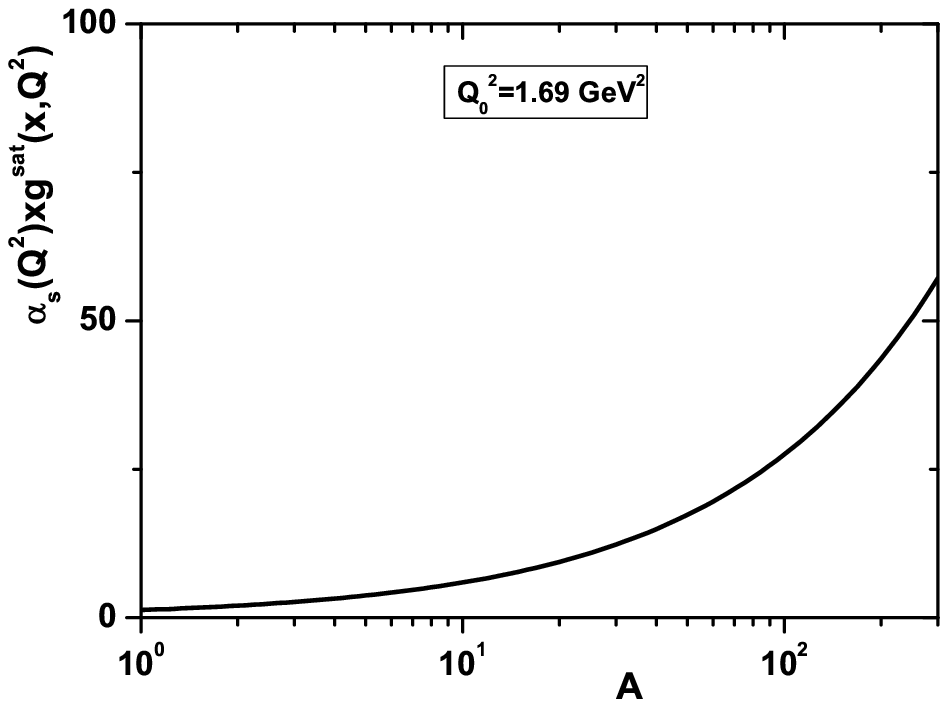}}
\caption{ Results of $\alpha_{s}xg^{A}_{\mathrm{sat}}$  in a wide
range of nuclei at $Q_{0}^2=1.69~\mathrm{GeV}^2$.}\label{Fig2}
\end{figure}
In Fig.1 we show the gluon saturation as a function of the mass
number A is expected to occur for various values of $Q^2$ [40]. In
Fig.2, the gluon distribution, $\alpha_{s}xg^{A}_{\mathrm{sat}}$
increases as the mass number A increase at the initial scale
$Q_{0}^{2}=1.69~\mathrm{GeV}^2$ [41]. Therefore, the effect of the
gluon saturation is expected to be larger in heavy nuclei, and to
be important for small values of $Q^2$. \\
We rewrite Eq.(17) by using Eq.(16), to take into account the
nonlinear correction to the nuclear gluon distribution at the
initial scale for $x<x_{0}$ as
\begin{eqnarray}
xg^{A,\mathrm{NLC}}(x,Q_{0}^2)=xg(x,Q_{0}^2)w_{g}(x,A,Z)\Big{\{}1+\frac{27{\pi}\alpha_{s}(Q_{0}^2)}{16\mathcal{R}^{2,A}Q_{0}^2}\Big{[}
xg(x,Q_{0}^2)w_{g}(x,A,Z)-xg(x_{0},Q_{0}^2)w_{g}(x_{0},A,Z)
\Big{]}\Big{\}}^{-1}.
\end{eqnarray}
We note that in Eq.(19), $xg^{A,\mathrm{NLC}}{\rightarrow}xg^{A}$
when $\mathcal{R}^{A}{\rightarrow}\infty$ and
$xg^{A}_{\mathrm{sat}}{\rightarrow}\infty$ and also we see that
$xg^{A,\mathrm{NLC}}{\rightarrow}xg^{A}_{\mathrm{sat}}$ when
$x{\rightarrow}0$. Moreover $xg^{A,\mathrm{NLC}}$ joins smoothly
onto $xg^{A}$ at $x=x_{0}(=10^{-2})$. The nonlinear corrections to
the gluon distribution are reflected in the sea-quark
distributions $q^{A}_{s}(x,Q^{2})$ which at small $x$ are
predominantly driven by the gluon and modified the nuclear
structure function, as\footnote{The shadowing corrections to the
gluon distribution are reflected in the seq-quark distributions
which the seq-quark starting distribution in the region $x<x_{0}$
in proportion to the shadowing correction to the gluon by the
following form [39]
$$
xq_{s}^{\mathrm{NLC}}(x,Q_{0}^2)=xq_{s}(x,Q_{0}^2)
\frac{xg^{\mathrm{NLC}}(x,Q_{0}^2)}{xg(x,Q_{0}^2)}.
$$ }
\begin{eqnarray}
xq_{s}^{A,\mathrm{NLC}}(x,Q_{0}^2)&=&xq_{s}^{A}(x,Q_{0}^2)
\frac{xg^{A,\mathrm{NLC}}(x,Q_{0}^2)}{xg^{A}(x,Q_{0}^2)}\nonumber\\
&&=xq_{s}^{A}(x,Q_{0}^2)\Big{\{}1+\theta(x_{0}-x)\Big{[}
xg^{A}(x,Q_{0}^2)-xg^{A}(x_{0},Q_{0}^2)
\Big{]}/xg^{A}_{\mathrm{sat}}(x,Q_{0}^{2})\Big{\}}^{-1}.
\end{eqnarray}
The weight function $w_{i}(x,A,Z)$ for the linear distribution
functions can be obtained from the three constrains for the
nuclear distributions as the nuclear charge Z, mass number A and
momentum conservations \footnote{The nonlinear terms will lead to
a very small violation of the momentum sum rules, which can be
recovered by a simple rescaling of the gluon distribution [39].
Recently, nonlinear corrections have been considered in Ref.[42]
for the nucleons and nuclei.} are defined by the following forms
[14,19, 33-38]
\begin{eqnarray}
Z&=&\int\frac{A}{3}\Big{[}2u_{v}^{A}-d_{v}^{A}\Big{]}(x,Q_{0}^{2})dx,\nonumber\\
A&=&\int\frac{A}{3}\Big{[}u_{v}^{A}+d_{v}^{A}\Big{]}(x,Q_{0}^{2})dx,\nonumber\\
A&=&\int
Ax\Big{[}u_{v}^{A}+d_{v}^{A}+2\{\overline{u}^{A}+\overline{d}^{A}
+\overline{s}^{A}\}+g^{A}\Big{]}(x,Q_{0}^{2})dx.
\end{eqnarray}
For a detailed investigation of these functions, we constrain our
results to the functions defined in Ref.[35]. The gluon
distribution at low $x$ is dominant, therefore we used the
standard gluon distribution at the input scale
$Q_{0}^{2}=1.69~\mathrm{GeV}^2$ obtained from CT18 set of the free
proton PDFs [43], i.e.,
\begin{eqnarray}
xg(x,Q_{0}^{2})=a_{0}x^{a_{1}}(1-x)^{a_{2}}\Big{[}\sinh(a_{3})(1-\sqrt{x})^3+
3\sinh(a_{4})\sqrt{x}(1-\sqrt{x})^2+(3+2a_{1})x(1-\sqrt{x})+x^{3/2}\Big{]},
\end{eqnarray}
where the coefficients $a_{0-4}$ are listed in Ref.[43]. The
weight function $w_{g}(x,A,Z)$ for the gluon distribution function
is defined by the following form [35]
\begin{eqnarray}
w_{g}(x,A,Z)&=&1+\Big{(}1-\frac{1}{A^{1/3}}\Big{)}(1-x)^{-\beta_{g}}\Big{[}
a_{g}(A)+xb_{g}(A)+x\Big{(}1-\frac{1}{A^{\epsilon_{bg}}}\Big{)}+
x^2c_{g}(A)\nonumber\\
&&+x^2\Big{(}1-\frac{1}{A^{\epsilon_{cg}}}\Big{)}+ x^3d_{g}(A)
\Big{]},
\end{eqnarray}
where the coefficients at the next-to-leading order (NLO) and the
next-to-next-to-leading order (NNLO) approximations are listed in
Ref.[35]. The strong coupling is set equal to
$\alpha_{s}(M_{z})=0.118$ for both the NLO and NNLO
approximations. In Figs.3 and 4, we show representations of the
nonlinear corrections to the gluon modification functions at the
initial scale $Q_{0}^{2}=1.69~\mathrm{GeV}^{2}$ for two selected
nuclei, C-12 and Pb-208 at the NLO and NNLO approximations,
respectively. The nuclear gluon distribution functions are
analyzed using the CT18 proton PDF set as a baseline in these
figures (i.e., Figs.3 and 4) [43]. The nuclear modification
factors have been extracted from QCD fits to the nuclear and
neutrino(antineutrino) DIS and Drell-Yan data\footnote {For future
discussions see Ref.[35].}.
\begin{figure}
\centerline{
\includegraphics[width=0.8\textwidth]{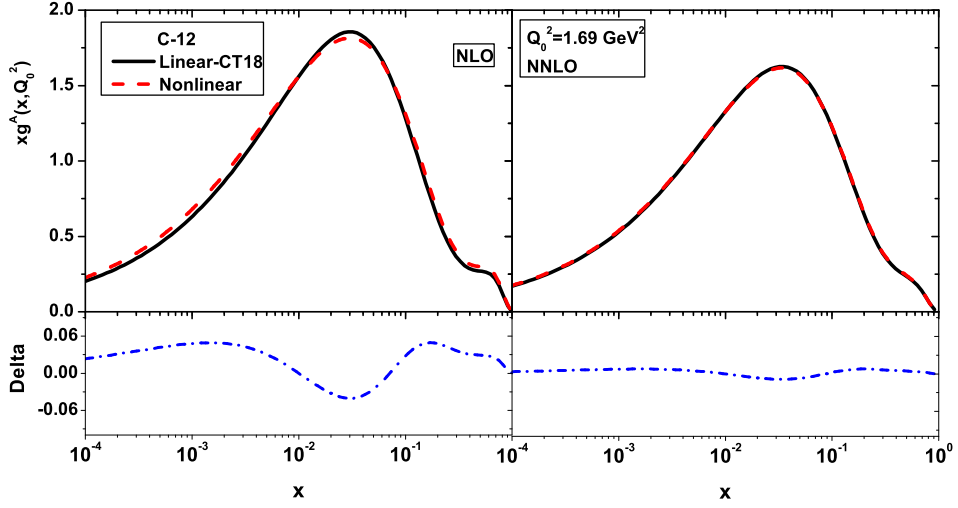}}
\caption{ The nonlinear gluon distribution function
($xg^{A,\mathrm{NLC}}(x,Q_{0}^{2})$) compared with the linear
($xg^{A}(x,Q_{0}^{2})$) for C-12 at $Q_{0}^2=1.69~\mathrm{GeV}^2$
in the NLO and NNLO approximations. The delta values are
differences between the nonlinear and linear distribution
functions
($\mathrm{Delta}=xg^{\mathrm{A,NLC}}(x,Q_{0}^{2})-xg^{A}(x,Q_{0}^{2})$)
at the initial scale. }\label{Fig3}
\end{figure}
\begin{figure}
\centerline{
\includegraphics[width=0.8\textwidth]{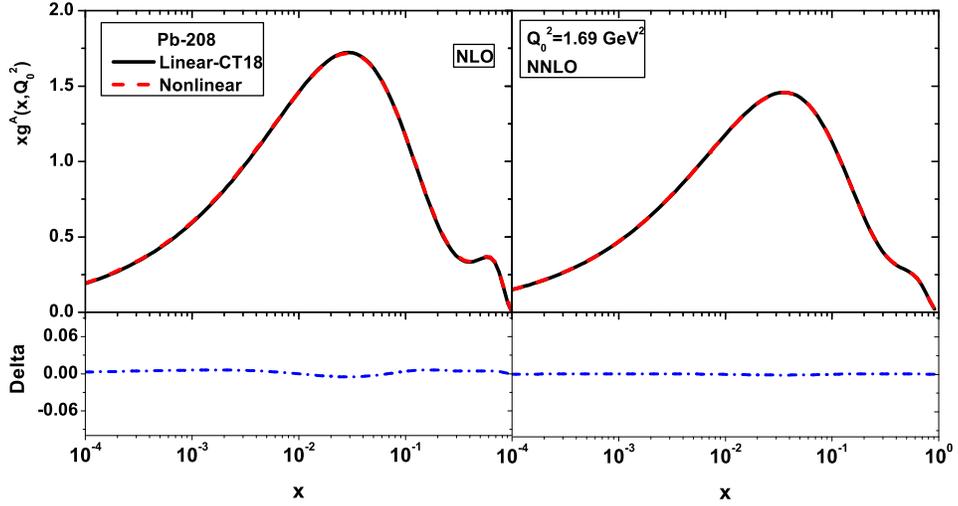}}
\caption{ The same as Fig.3 for lead.}\label{Fig4}
\end{figure}
The resulting nonlinear corrections to the gluon distribution
function are presented in Fig.5 for carbon (left) and iron (right)
at $Q_{0}^{2}=2~\mathrm{GeV}^{2}$ in the NLO approximation. To
achieve this, we used of the gluon distribution for a free proton
defined in Ref.[44] as
\begin{eqnarray}
xg(x,Q_{0}^{2})=A_{g}x^{\alpha_{g}}(1-x)^{\beta_{g}}(1
+\gamma_{g}x^{\delta_{g}}+\eta_{g}x),
\end{eqnarray}
where the coefficients at the NLO approximation are listed in
Refs.[19,44]. The weight function  for the nuclei of carbon and
iron has the same form in Eq. (16) which the coefficients are
presented in Ref.[19] in which the effects of shadowing,
anti-shadowing, fermi motion and the EMC regions are included.\\
\begin{figure}
\centerline{
\includegraphics[width=0.8\textwidth]{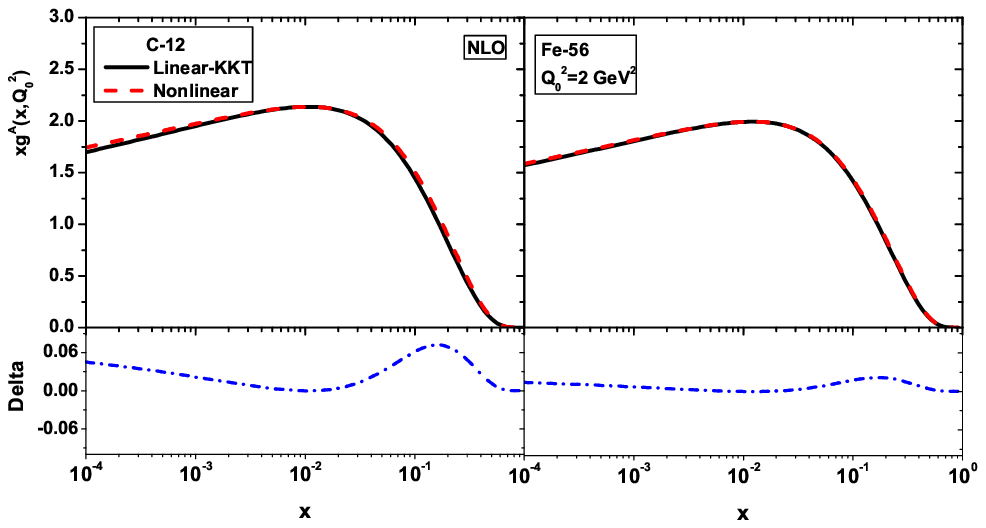}}
\caption{ The nonlinear gluon distribution function
($xg^{A,\mathrm{NLC}}(x,Q_{0}^{2})$) compared with the linear
($xg^{A}(x,Q_{0}^{2})$) for carbon (left) and iron (right) at
$Q_{0}^2=2~\mathrm{GeV}^2$ in the NLO approximation. The delta
values are differences between the nonlinear and linear
distribution functions
($\mathrm{Delta}=xg^{\mathrm{A,NLC}}(x,Q_{0}^{2})-xg^{A}(x,Q_{0}^{2})$)
at the initial scale.}\label{Fig5}
\end{figure}
In Fig.6, we compare the nonlinear and linear gluon distributions
in lead at the NNLO approximation to those of JR09 [45] at
$Q_{0}^{2}=2~\mathrm{GeV}^2$. The nuclear gluon distribution is
obtained from JR09 parametrization at the input scale by the
following form of the free proton PDFs
\begin{eqnarray}
xg(x,Q_{0}^{2})=3.0076x^{0.0637}(1-x)^{5.54473},
\end{eqnarray}
where the parameters in weight function is listed in Ref.[34]. To
quantify the magnitude of NNLO corrections, we present the
nonlinear corrections of nuclear gluon distributions obtained at
the input scale of the CT18 and JR09 parametrizations in Figs.3, 4
and 6 for light and heavy nuclei. The Delta functions in these
figures (i.e., Fis.3,4 and 6) show that the nonlinear and linear
gluon distributions have a similar behavior at the input scale in
a wide range of $x$. Therefore, Eq.(13) changes to an approximate
relation at the NNLO accuracy as
\begin{eqnarray}
xg^{A,NLC}(x,Q^{2})|_{\mathrm{NNLO}}&{\simeq}&xg^{A}(x,Q^{2})-\int_{Q_{0}^{2}}^{Q^2}\frac{81}{16}\frac{\alpha_{s}^{2}(Q^{2})}
{\mathcal{R}^{2,A}Q^{2}}d{\ln}Q^2\int_{\chi}^{1}\frac{dz}{z}[\frac{x}{z}g^{A}(\frac{x}{z},Q^{2})]^{2}\nonumber\\
&&=w_{g}(x,A,Z)xg(x,Q^{2})-\int_{Q_{0}^{2}}^{Q^2}\frac{81}{16}\frac{\alpha_{s}^{2}(Q^{2})}
{\mathcal{R}^{2,A}Q^{2}}d{\ln}Q^2\int_{\chi}^{1}\frac{dz}{z}w_{g}^{2}(\frac{x}{z},A,Z)[\frac{x}{z}g(\frac{x}{z},Q^{2})]^{2}
\end{eqnarray}
In Figs.3-5, we observe that
$xg^{A,NLC}(x,Q_{0}^{2}){\neq}xg^{A}(x,Q_{0}^{2})$ at the NLO
accuracy, therefore the evolution of the nuclear gluon
distribution functions with the nonlinear corrections are defined
by the following form
\begin{eqnarray}
xg^{A,NLC}(x,Q^{2})|_{\mathrm{NLO}}&=&xg(x,Q_{0}^2)w_{g}(x,A,Z)\Big{[}\Big{\{}1+\frac{27{\pi}\alpha_{s}(Q_{0}^2)}{16\mathcal{R}^{2,A}Q_{0}^2}\Big{[}
xg(x,Q_{0}^2)w_{g}(x,A,Z)-xg(x_{0},Q_{0}^2)w_{g}(x_{0},A,Z)
\Big{]}\Big{\}}^{-1}-1\Big{]}\nonumber\\
&&+w_{g}(x,A,Z)xg(x,Q^{2})-\int_{Q_{0}^{2}}^{Q^2}\frac{81}{16}\frac{\alpha_{s}^{2}(Q^{2})}
{\mathcal{R}^{2,A}Q^{2}}d{\ln}Q^2\int_{\chi}^{1}\frac{dz}{z}w_{g}^{2}(\frac{x}{z},A,Z)[\frac{x}{z}g(\frac{x}{z},Q^{2})]^{2}.
\end{eqnarray}
\begin{figure}
\centerline{
\includegraphics[width=0.5\textwidth]{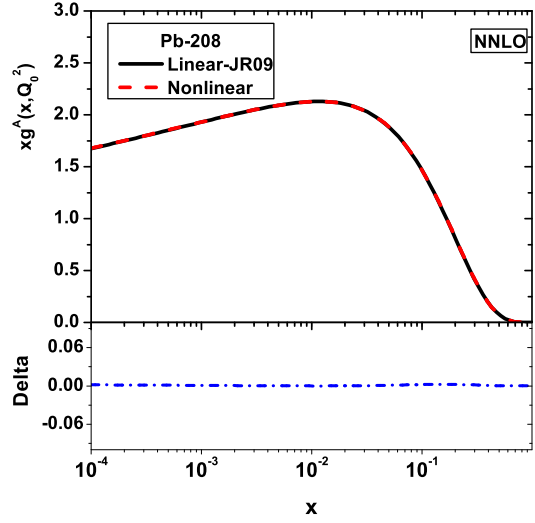}}
\caption{ The nonlinear gluon distribution function
($xg^{A,\mathrm{NLC}}(x,Q_{0}^{2})$) compared with the linear
($xg^{A}(x,Q_{0}^{2})$) for lead at $Q_{0}^2=2~\mathrm{GeV}^2$ in
the NNLO approximation. The delta values are differences between
the nonlinear and linear distribution functions
($\mathrm{Delta}=xg^{\mathrm{A,NLC}}(x,Q_{0}^{2})-xg^{A}(x,Q_{0}^{2})$)
at the initial scale.}\label{Fig6}
\end{figure}
For evolution of the nonlinear corrections of the nuclear gluon
distributions, we need to a gluon analytical distribution function
for a free proton at the scale $Q^2$ in the NLO and NNLO
approximations. In literatures, usually, the gluon analytical
distribution function at the LO approximation have been defined.
To do it, we extend the analytical solution used in the DGLAP
evolution  to take into account the nonlinear corrections in
Eqs.(27) and (26) at the NLO and NNLO approximations,
respectively. We solve the DGLAP evolution equation using Laplace
transform techniques in the next section.\\

\subsection{3. Higher order corrections to the gluon distribution}

According to the DGLAP $Q^{2}$-evolution equations, the
 singlet distribution function leads to the following relation of integro-differential equation
\begin{eqnarray}
\frac{{\partial}F_{2}(x,Q^{2})}{{\partial}{\ln}Q^{2}}&=&P_{qq}(x){\otimes}F_{2}(x,Q^{2})+<e^{2}>P_{qg}(x){\otimes}xg(x,Q^{2})
\end{eqnarray}
where $P_{qq}$ and $P_{qg}$ are the quark-quark and quark-gluon
splitting functions calculated to the desired order in
$\alpha_{s}$ [46-48]. Here $<e^{2}>$ is the average of the charge
$e^{2}$ for the active quark flavors. Also,
$<e^{2}>=n_{f}^{-1}\sum_{i=1}^{n_{f}}e_{i}^{2}$, and the symbol
$\otimes$ denotes convolution according to the usual prescription.
Considering the variable definitions $\upsilon{\equiv}\ln(1/x)$
and $w{\equiv}\ln(1/z)$, one can rewrite Eq. (28) in terms of the
convolution integrals and new variables as
\begin{eqnarray}
\frac{\partial{\mathcal{\widehat{F}}_{2}(\upsilon,Q^{2})}}{\partial{\ln}Q^{2}}&=&\int_{0}^{\upsilon}[\mathcal{\widehat{F}}_{2}(\upsilon,Q^{2})
\mathcal{\widehat{H}}^{(\varphi)}_{2,s}(\alpha_{s}(Q^{2}),\upsilon-w)
+<e^{2}>\mathcal{\widehat{G}}(\upsilon,Q^{2})
\mathcal{\widehat{H}}^{(\varphi)}_{2,g}(\alpha_{s}(Q^{2}),\upsilon-w)]dw,
\end{eqnarray}
where
\begin{eqnarray}
\frac{\partial{\mathcal{\widehat{F}}_{2}(\upsilon,Q^{2})}}{\partial{\ln}Q^{2}}&{\equiv}&
\frac{{\partial}F_{2}(e^{-\upsilon},Q^{2})}{\partial{\ln}Q^{2}},\nonumber\\
\mathcal{\widehat{G}}(\upsilon,Q^{2})&{\equiv}&G(e^{-\upsilon},Q^{2}),\nonumber\\
\mathcal{\widehat{H}}^{(\varphi)}(\alpha_{s}(Q^{2}),\upsilon)&{\equiv}&e^{-\upsilon}\widehat{P}_{a,b}^{(\varphi)}(\alpha_{s}(Q^{2}),\upsilon),\nonumber\\
\end{eqnarray}
The Laplace transform of
$\mathcal{\widehat{H}}(\alpha_{s}(Q^{2}),\upsilon)$$^{,}s$
 are given by the following forms
\begin{eqnarray}
\Phi_{f}^{(\varphi)}(\alpha_{s}(Q^{2}),s)&{\equiv}&
{\mathcal{L}}[\mathcal{\widehat{H}}^{(\varphi)}_{2,s}(\alpha_{s}(Q^{2}),\upsilon);s]=\int_{0}^{\infty}\mathcal{\widehat{H}}^{(\varphi)}_{2,s}(\alpha_{s}(Q^{2}),\upsilon)e^{-s\upsilon}d\upsilon,\nonumber\\
 \Theta_{f}^{(\varphi)}(\alpha_{s}(Q^{2}),s)&{\equiv}&{\mathcal{L}}[\mathcal{\widehat{H}}^{(\varphi)}_{2,g}(\alpha_{s}(Q^{2}),\upsilon);s]=\int_{0}^{\infty}\mathcal{\widehat{H}}^{(\varphi)}_{2,g}(a_{s}(Q^{2}),\upsilon)e^{-s\upsilon}d\upsilon.\nonumber\\
\end{eqnarray}
Consequently, we can  rewrite Eq.(29) in the Laplace space $s$, by
using the convolution theorem for Laplace transforms and
considering the fact that the Laplace transform of the convolution
factors are simply the ordinary product of the Laplace transform
of the factors, i.e.,
\begin{eqnarray}
\frac{\partial{f_{2}(s,Q^{2})}}{\partial{\ln}Q^{2}}&=&
\Phi_{f}^{(\varphi)}(\alpha_{s}(Q^{2}),s)f_{2}(s,Q^{2})+<e^{2}>\Theta_{f}^{(\varphi)}(\alpha_{s}(Q^{2}),s)g(s,Q^{2}),
\end{eqnarray}
where
\begin{eqnarray}
{\mathcal{L}}[\mathcal{\widehat{F}}_{2}(\upsilon,Q^{2});s]&=&f_{2}(s,Q^{2}),
\end{eqnarray}
and
\begin{eqnarray}
\eta_{f}^{(\varphi)}(\alpha_{s}(Q^{2}),s)&=&\sum_{\phi=0}^{\varphi}\alpha_{s}^{\phi+1}
(Q^{2})\eta^{(\phi)}_{f}(s),
~~~~~~~\mathrm{for}~~~~~~~~~\eta=(\Phi, \Theta ),
\end{eqnarray}
The coefficient functions $\Phi$ and $\Theta$ in the Laplace space
$s$ at the LO approximation are given by
\begin{eqnarray}
\Theta_{f}^{(0)}(s)&=&2n_{f}(\frac{1}{1+s}-\frac{2}{2+s}+\frac{2}{3+s}),\\
\Phi_{f}^{(0)}(s)&=&4-\frac{8}{3}(\frac{1}{1+s}+\frac{1}{2+s}+2(\psi(s+1)+\gamma_{E})),
\end{eqnarray}
where $\psi(x)$ is the digamma function and $\gamma_{E}=0.5772156
. . .$ is Euler constant.\\
The explicit expressions for the NLO and NNLO kernels in $s$ space
are rather cumbersome; therefore, we recall that we are interested
in investigation of the kernels in small $x$ [49-51]. In the
Laplace space, we consider the kernels at small $s$, as the two
and three-loop kernels read
\begin{eqnarray}
\Theta_{f,s{\rightarrow}0}^{(1)}(s)&{\simeq}&C_{A}T_{f}[\frac{40}{9s}],\nonumber\\
\Phi_{f,s{\rightarrow}0}^{(1)}(s)&{\simeq}&C_{F}T_{f}[\frac{40}{9s}],
\end{eqnarray}
and
\begin{eqnarray}
\Theta_{f,s{\rightarrow}0}^{(2)}(s)&{\simeq}&n_{f}[-\frac{1268.300}{s}+\frac{896}{3s^{2}}]+n^{2}_{f}[\frac{1112}{243s}],\nonumber\\
\Phi_{f,s{\rightarrow}0}^{(2)}(s)&{\simeq}&n_{f}[-\frac{506}{s}+\frac{3584}{27s^{2}}]+n^{2}_{f}[\frac{256}{81s}],
\end{eqnarray}
with the color factors $C_{A}=N_{c}=3$,
$C_{F}=\frac{N_{c}^{2}-1}{2N_{c}}=\frac{4}{3}$ and
$T_{f}=\frac{1}{2}n_{f}$ associated with the color group $SU(3)$
and $n_{f}$ being the
number of flavors.\\
The strong coupling satisfies the renormalization group equation,
which up to NNLO reads
$$
\frac{d}{d\ln{Q^2}}\bigg{(}\frac{\alpha_{s}}{4\pi}\bigg{)}=-\beta_{0}\bigg{(}\frac{\alpha_{s}}{4\pi}\bigg{)}^2-
\beta_{1}\bigg{(}\frac{\alpha_{s}}{4\pi}\bigg{)}^3
-\beta_{2}\bigg{(}\frac{\alpha_{s}}{4\pi}\bigg{)}^4-...
$$
where $\beta_{0}$, $\beta_{1}$ and $\beta_{2}$ are the one, two
and three loop correction to the QCD $\beta$-function. The
standard representation for QCD couplings in NLO and NNLO (within
the $\mathrm{\overline{MS}}$-scheme) approximations have the forms
\begin{eqnarray}
\alpha_{s}(t)&=&\frac{4\pi}{\beta_{0}t}\Big{[}1
-\frac{\beta_{1}}{\beta_{0}^{2}}\frac{\ln{t}}{t}\Big{]}~~~~~~~~~~~~~~~~~~~~~~~~~~~~~(\mathrm{NLO}),\nonumber\\
\alpha_{s}(t)&=&\frac{4\pi}{\beta_{0}t}\Big{[}1
-\frac{\beta_{1}}{\beta_{0}^{2}}\frac{\ln{t}}{t}+\frac{1}{\beta_{0}^{3}t^{2}}\bigg{\{}
\frac{\beta_{1}^{2}}{\beta_{0}}(\ln^{2}t-\ln{t}-1)+\beta_{2}\bigg{\}}
\Big{]}~~~(\mathrm{NNLO}),
\end{eqnarray}
where $t=\ln\frac{Q^{2}}{\Lambda^{2}}$ and $\Lambda$ is the QCD
cut-off
parameter [52].\\
Consequently, the discretized form of Eq.(32) for the gluon
distribution reads
\begin{eqnarray}
g(s,Q^{2})&=&h^{(\varphi)}(\alpha_{s}(Q^{2}),s)\frac{\partial{f_{2}(s,Q^{2})}}{\partial{\ln}Q^{2}}-k^{(\varphi)}(\alpha_{s}(Q^{2}),s)f_{2}(s,Q^{2}),
\end{eqnarray}
where the kernels $k^{(\varphi)}(\alpha_{s}(Q^{2}),s)$ and
$h^{(\varphi)}(\alpha_{s}(Q^{2}),s)$ contain contributions of the
$s$-space splitting and coefficient functions up to the NNLO
approximation. These kernels can be evaluated from $s$-space
results by the following forms
\begin{eqnarray}
h^{(\varphi)}(\alpha_{s}(Q^{2}),s)&=&\frac{1}{<e^2>\sum_{\phi=0}^{\varphi}\alpha_{s}^{\phi+1}(Q^{2})\Theta^{(\phi)}_{f}(s)},\nonumber\\
k^{(\varphi)}(\alpha_{s}(Q^{2}),s)&=&\frac{\sum_{\phi=0}^{\varphi}\alpha_{s}^{\phi+1}(Q^{2})\Phi^{(\phi)}_{f}(s)}{<e^2>\sum_{\phi=0}^{\varphi}\alpha_{s}^{\phi+1}(Q^{2})\Theta^{(\phi)}_{f}(s)}.
\end{eqnarray}
The inverse Laplace transform of coefficients $k(a_{s}(Q^{2}),s)$
and $h(a_{s}(Q^{2}),s)$ in above equations are defined
respectively as kernels
$$\widehat{\eta}(a_{s}(Q^{2}),\upsilon){\equiv}{\mathcal{L}}^{-1}[k(\alpha_{s}(Q^{2}),s);\upsilon]$$
and
$$\widehat{J}(a_{s}(Q^{2}),\upsilon){\equiv}{\mathcal{L}}^{-1}[h(\alpha_{s}(Q^{2}),s);\upsilon].$$
The kernels  are dependent on $\upsilon$ and the running coupling
at the higher order approximations. In order to obtain an
analytical form for these kernels at higher order approximations,
we consider  the terms of the order $1/s$ as these terms are
dominant at higher order [53]. Therefore, we have
\begin{eqnarray}
\widehat{g}(\upsilon,Q^{2})&{\equiv}&{\mathcal{L}}^{-1}[g(s,Q^{2});\upsilon]=\int_{0}^{\upsilon}[\frac{\partial{\widehat{F}_{2}(w,Q^{2})}}
{\partial{\ln}Q^{2}}\widehat{J}^{(\varphi)}(\alpha_{s}(Q^{2}),\upsilon-w)-\widehat{F}_{2}(w,Q^{2})\widehat{\eta}^{(\varphi)}(\alpha_{s}(Q^{2}),\upsilon-w)]dw.\nonumber
\end{eqnarray}
Consequently, the general analytical expressions for the gluon
distribution function in $x$-space at the higher order
approximations are given by
\begin{eqnarray}
xg^{(\varphi)}(x,Q^{2})&=&\int_{x}^{1}\frac{dy}{y}[\frac{{\partial}F_{2}(y,Q^{2})}{\partial{\ln}Q^{2}}J^{(\varphi)}({\frac{x}{y}},Q^{2})-
F_{2}(y,Q^{2})\eta^{(\varphi)}({\frac{x}{y}},Q^{2})].
\end{eqnarray}
Having an analytical proton structure function and its derivative
with respect to $\ln{Q^{2}}$, one can extract the gluon
distribution function at any desired $x$ and
$Q^{2}$ values.\\
Using a parameterization suggested by authors in Ref.[54]  on the
proton structure function in a full accordance with the Froissart
predictions [55]. The explicit expression for the $F_{2}$
parameterization, obtained from a combined fit of the H1 and ZEUS
collaborations data [56] in the range of the kinematical variables
$x$ and $Q^{2}$( $x<0.01$ and $0.15< Q^{2}<3000~\mathrm{GeV}
^{2}$), is given by
\begin{eqnarray}
F_{ 2}(x,Q^{2})& =& D(Q^{2})(1-
x)^{n}\sum_{m=0}^{2}A_{m}(Q^{2})L^{m},
\end{eqnarray}
and
\begin{eqnarray}
\frac{{\partial}F_{2}(x,Q^{2})}{\partial{\ln}Q^{2}}&=&
F_{2}(x,Q^{2})[\frac{{\partial}{\ln}D(Q^{2})}{\partial{\ln}Q^{2}}+\frac{{\partial}{\ln}\sum_{m=0}^{2}A_{m}(Q^{2})L^{m}}{\partial{\ln}Q^{2}}],\nonumber
\end{eqnarray}
where
\begin{eqnarray}
A_{0}(Q^{2})&=&a_{00} + a_{01}
{\ln}(1+\frac{Q^{2}}{\mu^{2}}),~~~~~ A_{1}(Q^{2})=a_{10} + a_{11}
{\ln}(1+\frac{Q^{2}}{\mu^{2}}) +
a_{12}{\ln}^{2}(1+\frac{Q^{2}}{\mu^{2}})
 ,\nonumber\\
A_{2}(Q^{2})&=&a_{20} + a_{21} {\ln}(1+\frac{Q^{2}}{\mu^{2}}) +
a_{22}{\ln}^{2}(1+\frac{Q^{2}}{\mu^{2}})
 ,~~~D(Q^{2})=\frac{Q^{2}(Q^{2}+\lambda M^{2})}{(Q^{2}+M^{2})^2},~~~L^{m}=\ln^{m}(\frac{1}{x}\frac{Q^{2}}{Q^{2}+\mu^{2}}).
\end{eqnarray}
Here $M$ and $\mu^{2}$ are the effective mass  a scale factor
respectively. The effective parameters in Eq.(44) are defined in Refs.[54] and [57].\\
\begin{figure}
\centerline{
\includegraphics[width=0.65\textwidth]{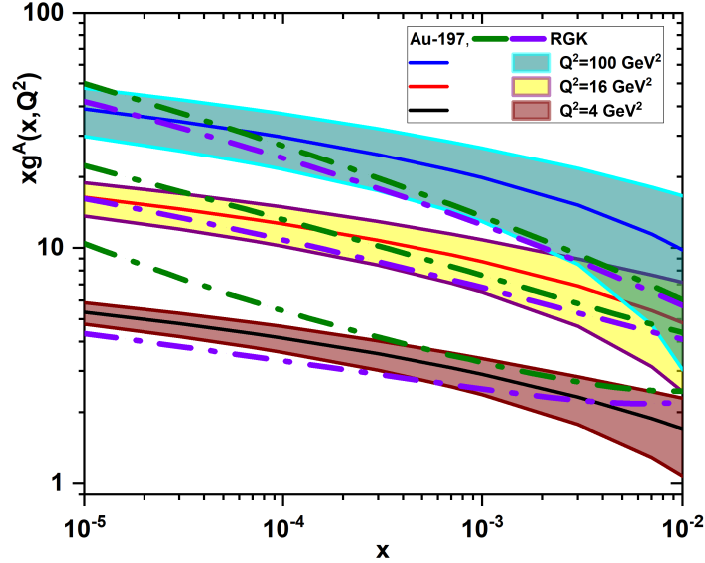}}
\caption{ The nonlinear gluon distribution function
($xg^{A,\mathrm{NLC}}(x,Q^{2})$) and their uncertainties at
$Q^2=4,~ 16$ and $100~\mathrm{GeV}^2$ for Au-197 compared with the
results of the nonlinear GLR-MQ gluon distribution function (the
RGK model) [33]. The dashed-dot lines (green and purple curves)
are upward and downward evolutions [33]. }\label{Fig7}
\end{figure}
\subsection{4. Results for nonlinear nuclear gluon distribution function}

Using the analytical approach outlined above (i.e., Eq.(42)) in
the NLO and NNLO approximations, we solve the nonlinear gluon
distributions for nuclei at low $x$ as
\begin{eqnarray}
xg^{A,NLC}(x,Q^{2})|_{\mathrm{NLO}}&=&xg(x,Q_{0}^2)w_{g}(x,A,Z)\Big{[}\Big{\{}1+\frac{27{\pi}\alpha_{s}(Q_{0}^2)}{16\mathcal{R}^{2,A}Q_{0}^2}\Big{[}
xg(x,Q_{0}^2)w_{g}(x,A,Z)-xg(x_{0},Q_{0}^2)w_{g}(x_{0},A,Z)
\Big{]}\Big{\}}^{-1}-1\Big{]}\nonumber\\
&&+w_{g}(x,A,Z)xg^{(1)}(x,Q^{2})-\int_{Q_{0}^{2}}^{Q^2}\frac{81}{16}\frac{\alpha_{s}^{2}(Q^{2})}
{\mathcal{R}^{2,A}Q^{2}}d{\ln}Q^2\int_{\chi}^{1}\frac{dz}{z}w_{g}^{2}(\frac{x}{z},A,Z)[\frac{x}{z}g^{(1)}(\frac{x}{z},Q^{2})]^{2},
\end{eqnarray}
and
\begin{eqnarray}
xg^{A,NLC}(x,Q^{2})|_{\mathrm{NNLO}}&{\simeq}&w_{g}(x,A,Z)xg^{(2)}(x,Q^{2})-\int_{Q_{0}^{2}}^{Q^2}\frac{81}{16}\frac{\alpha_{s}^{2}(Q^{2})}
{\mathcal{R}^{2,A}Q^{2}}d{\ln}Q^2\int_{\chi}^{1}\frac{dz}{z}w_{g}^{2}(\frac{x}{z},A,Z)[\frac{x}{z}g^{(2)}(\frac{x}{z},Q^{2})]^{2}.
\end{eqnarray}
Now we present our numerical results of the nonlinear gluon
distribution for light and heavy nuclei in the $x-Q^2$ kinematic
regions, where the nonlinear corrections are important. The
computed results of the nonlinear gluon distribution  function for
Au-197  compare with the suggested method by Rausch, Guzey and
Klasen (the RGK model)[33]. This was based on the brute force
method, where the authors in Ref.[33] have been extended the
numerical algorithm used in the $\mathrm{QCDNUM16}$  DGLAP
evolution code [58] to take into account the nonlinear corrections
as the nCTEQ15 nPDFs [59]
are used as baseline PDFs.\\
In Fig.7 we show representations of the nonlinear gluon
distribution functions for Au-197 at the scales $Q^2=4,~ 16$ and
$100~\mathrm{GeV}^2$ as a function of the momentum fraction $x$ to
show the effects of the $Q^2$ evolution. The nuclear weight
functions for the gluon are extracted from the suggested method by
Khanpour, Soleymaninia, Atashbar Tehrani, Spiesberger and Guzey
(the KSASG20 model) [35], where the CT18 nPDFs [43] are used as
baseline PDFs. These results are compared to the RGK model [33],
where the nCTEQ15 nPDFs [59] are used as baseline PDFs in the
nonlinear GLR-MQ evolution equation. In the RGK model, the
dashed-dot curves show the results of the upward evolution from
$Q_{0}^{2}=4~\mathrm{GeV}^2$ to $Q^2=16$ and $100~\mathrm{GeV}^2$
(green curves) and also show the results of the downward evolution
from $Q_{0}^{2}=100~\mathrm{GeV}^2$ to $Q^2=16$ and
$4~\mathrm{GeV}^2$ (purple curves) [33], respectively. The
uncertainties, due to the statistical errors of the coefficient
functions of the parametrization of the proton structure function
[54] and the nuclear modification functions [35], are shown in
Fig.7. For the NLO analysis, the nonlinear nuclear distribution
function for the gluon shows an increase as $x$ decreases, which
is similar to what one can observe in the analyses by RGK [33].
However, the magnitude of these results is slightly differs at
different scales, but they are within the uncertainties error
bands. As can be seen in the figure, the nonlinear gluon densities
come with relatively large error bands at the critical point
between the linear and nonlinear (i.e., $x=0.01$), reflecting the
fact that there are large errors due to the coefficients in the
parametrization of the proton structure function.\\
\begin{figure}
\centerline{
\includegraphics[width=0.65\textwidth]{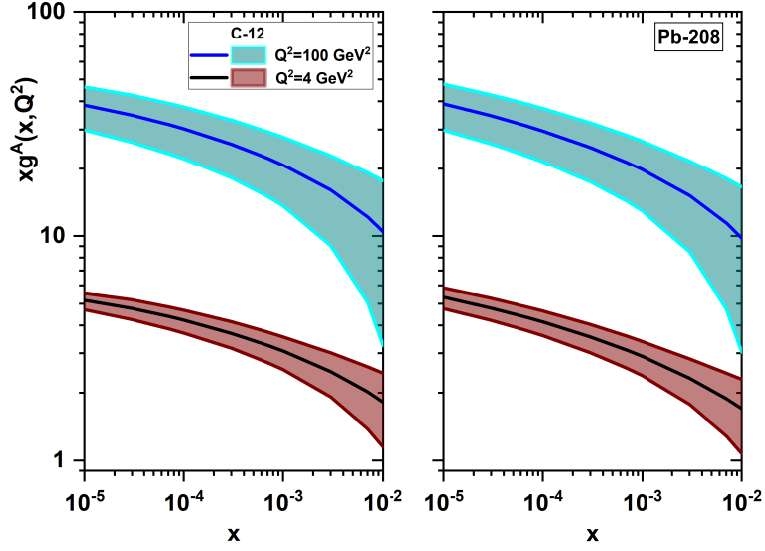}}
\caption{ The nonlinear gluon distribution function
($xg^{A,\mathrm{NLC}}(x,Q^{2})$) and their uncertainties at
$Q^2=4$ and $100~\mathrm{GeV}^2$ for C-12(left) and Pb-208(right)
as a function of $x$.}\label{Fig8}
\end{figure}
In Fig.8, the  nonlinear gluon distributions for C-12 and Pb-208
at the NLO approximation are considered at $Q^2=4$ and
$100~\mathrm{GeV}^2$ as a function of $x$ as accompanied with
their uncertainties. To quantify the magnitude of the nonlinear
corrections, we present ratios of nuclear gluon distributions
obtained in the nonlinear corrections over those of the linear.
Figure 9 quantifies the size of the nonlinear corrections as a
function of the mass number A and $x$ for C-12 and Au-197 at
$Q^2=4$ and $100~\mathrm{GeV}^2$. The difference between the
nonlinear and linear evolved gluon densities grows steadily with a
decrease of $x$. This is largest at the smallest values of $x$ and
$Q^2$ and disappears for $x=0.01$. The saturation gluon increases
as the atomic number increases, therefore the nonlinear/linear
ratio decreases as the atomic number increases. As one can see,
the nonlinear/linear ratio is slightly larger for light nuclei
than for heavy nuclei, and this effect is, as expected, mainly due
to the large gluon saturation values of heavy nuclei.\\
\begin{figure}
\centerline{
\includegraphics[width=0.65\textwidth]{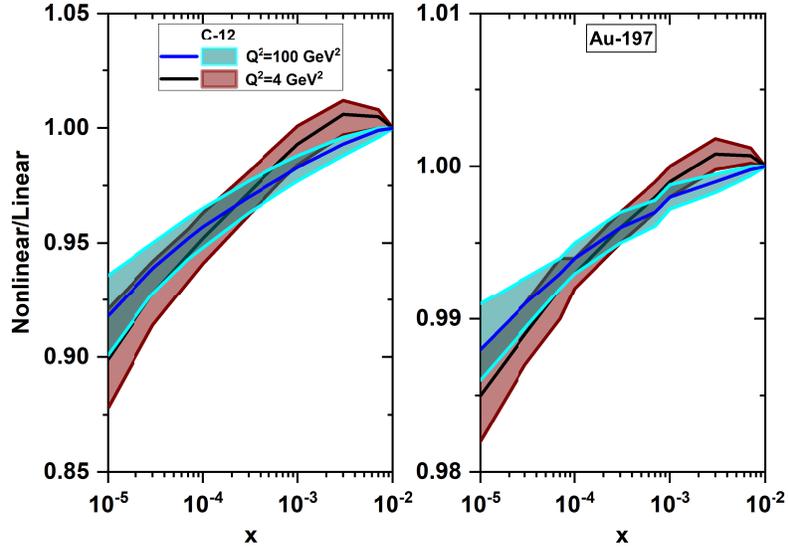}}
\caption{ The ratio of the nonlinear/linear gluon distributions
for C-12 and Au-197 at $Q^2=4$ and $100~\mathrm{GeV}^2$ as a
function of $x$ with their uncertainties.}\label{Fig9}
\end{figure}
\begin{figure}
\centerline{
\includegraphics[width=0.65\textwidth]{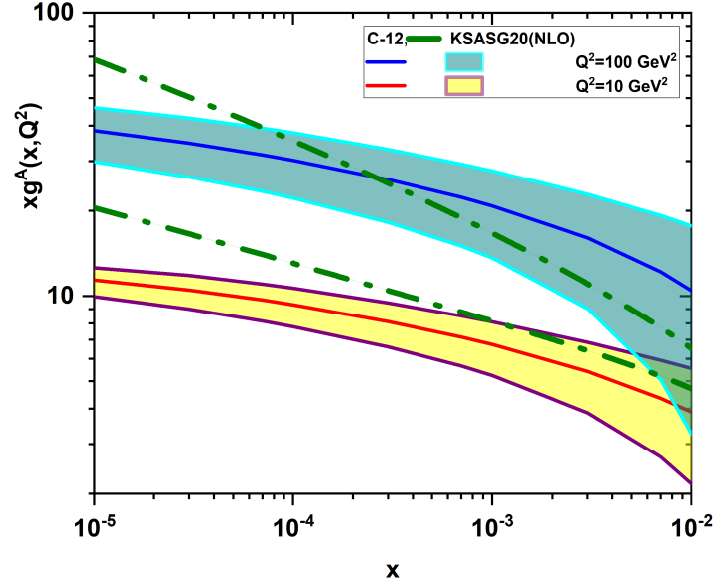}}
\caption{ The nonlinear gluon distribution function
($xg^{A,\mathrm{NLC}}(x,Q^{2})$) and their uncertainties at
$Q^2=10$ and $100~\mathrm{GeV}^2$ for C-12 compared with the
linear results based on the  KSASG20 (NLO) model (dashed-dot
curves) [35]. }\label{Fig10}
\end{figure}\begin{figure}
\centerline{
\includegraphics[width=0.65\textwidth]{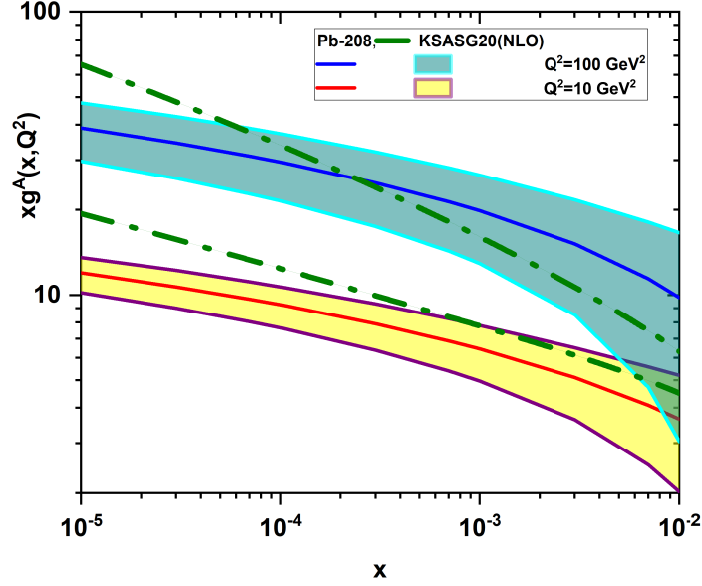}}
\caption{ The same as Fig.10 for Pb-208. }\label{Fig11}
\end{figure}
In Figs. (10) and (11), the nonlinear corrections to the gluon
distribution function at the NLO approximation for the nuclei C-12
and Pb-208 are presented at $Q^2=10$ and $100~\mathrm{GeV}^2$ as a
function of the momentum fraction $x$, respectively. In these
figures, our numerical results, which are accompanied with
 statistical errors, are compared with the linear results based on
the KSASG20 (NLO) parametrization [35]. The KSASG20
parametrization is a new set of nuclear parton distribution
functions (nuclear PDFs) at the NLO and NNLO approximations in
perturbative QCD which include the new CT18 PDFs on proton PDFs.
 As can be seen in these
figures, the effects of nonlinear corrections are noticeable at
small $x$ values and the strong growth of gluon distributions are
tamed by shadowing effects as $x$ decreases. The solid curves
represent the effect of shadowing correction for
$\mathcal{R}^{A}=2A^{1/3}~\mathrm{GeV}^{-1}$ presented by using
Eq.(45). As can be observed, the nuclear gluon distributions
increase as $x$ decreases, which corresponds with the perturbative
QCD fits at small $x$, but these behaviors are tamed with respect
to nonlinear terms at the GLR-MQ equation. These tamed behaviors
of nuclear gluon distributions due to the shadowing corrections
satisfy the Froissart bound in the perturbative QCD means. Hence,
as one can see from Figs.10 and 11, deviations from the linear
nuclear gluon distributions based on the KSASG20 (NLO)
parametrization increases as $x$ decreases. The deviations from
the KSASG20 (NLO) nPDFs increases as $Q^2$ increases and decreases
as atomic number $A$  increases (indeed the nonlinear nuclear
gluon distributions increases as atomic number $A$ increases), and
significant effects are found for heavier nuclei, such as lead.
These behaviors for the nonlinear nuclear gluon distributions are similar to the analysis of RGK [33].\\

\subsection{5. Summary}

In conclusion, we have studied the effects of adding the nonlinear
corrections to the gluon distribution function for light and heavy
nuclei at small $x$ analytically. We used the parametrization of
the proton structure function to take into account an analytical
solution for the gluon density at low $x$ in the NLO
approximation. The nuclear modification factors are obtained from
KSASG20 nuclear PDFS which are based on the CT18 framework. The
shadowing effects of the gluon distribution at small $x$ through
modifications of the starting distributions and the presence of
additional nonlinear terms in the initial point $Q_{0}^{2}$ at the
NLO and NNLO approximations for light and heavy nuclei considered.
We obtained the nonlinear corrections for small $x$ in a wide
range of $Q^2$ values. These results show that the magnitude of
the nonlinear corrections increases with a decrease of $x$ and an
increase of the atomic number $A$. Our results are consistent,
within uncertainties, with the determination of nuclear gluon
distribution with the upward and downward evolution from the RGK
model, which is based on the nCTEQ15 nPDFs as input. Our
determination of nuclear gluon distributions includes error
estimates obtained with respect to the coefficient errors in the
parametrization of the proton structure function and the nuclear
modification function errors. We found differences between our and
RGK results at the NLO accuracy, which occur in  different
assumptions such as the input prameterizations and also the
approximate relation between the gluon distribution and the proton
structure function due to the Laplace transform method at low $x$.
These results for the nonlinear corrections to the nuclear gluon
distribution function may be important for future experiments at
the Electron-Ion Collider [21,22], LHeC Collaboration or a Future
Circular Collider (FCC) study group [5] and Electron-Ion
Collider in China (EiCC) [60] at low $x$.\\



\subsection{ACKNOWLEDGMENTS}

We are grateful to the Razi University for financial support of
this project. G.R.Boroun thanks V. Guzey for allowing access to
data related to the nonlinear corrections for the gluon
distribution function
 for Au-197.\\

\subsection{Appendix}
Previous studies of the GLR-MQ terms in the context of extracting
the parton distribution functions can be found in Ref.[39]. The
nonlinear evolution equations relevant at high gluon densities
have been studied at small $x$ where we expect annihilation or
recombination of gluons to occur. A measurement of $g(x,Q^2)$ in
this region probes a gluon of transverse size $\sim 1/Q$,
therefore the transverse area of the thin disc that they occupy is
$\sim xg(x,Q^2)/Q^2$. The shadowing effects, at sufficiently small
$x$ where $W{\lesssim}\alpha_{s}$, can be calculated in
perturbative QCD. Here
$W{\sim}\frac{\alpha_{s}(Q^2)}{\pi{\mathcal{R}^2}Q^2}xg(x,Q^2)$
where $\pi{\mathcal{R}^2}$ is the transverse area and
$\mathcal{R}$ is the proton radius. The QCD evolution equation
modified for the gluon distribution is defined by the following
form
\begin{eqnarray}
\frac{\partial{xg(x,Q^{2})}}{\partial{\ln}Q^{2}}=P_{gg}{\otimes}xg+P_{gq}{\otimes}xq_{s}
-\frac{81}{16}\frac{\alpha_{s}^{2}(Q^{2})}{\mathcal{R}^{2}Q^{2}}\theta(x_{0}-x)
\int_{x}^{x_{0}}\frac{dx'}{x'}[{x'}g({x'},Q^{2})]^{2},
\end{eqnarray}
where the $\theta$ function reflects the ordering in longitudinal
momenta as for $x{\geq}x_{0}$ the shadowing correction is
negligible ($x_{0}{=}10^{-2}$).  The shadowing term has a minus
sign  because the scattering amplitude corresponding to the gluon
ladder is predominantly imaginary. Equation (47) can be rewritten
with a variable change ($x'=\frac{x}{z}$) as
\begin{eqnarray}
\frac{\partial{xg(x,Q^{2})}}{\partial{\ln}Q^{2}}=\frac{\partial{xg(x,Q^{2})}}{\partial{\ln}Q^{2}}|_{\mathrm{DGLAP}}
-\frac{81}{16}\frac{\alpha_{s}^{2}(Q^{2})}{\mathcal{R}^{2}Q^{2}}\theta(x_{0}-x)
\int_{\chi}^{1}\frac{dz}{z}[\frac{x}{z}g(\frac{x}{z},Q^{2})]^{2}.
\end{eqnarray}
There are also shadowing corrections to the evolution equation for
the sea-quark distributions as
\begin{eqnarray}
\frac{\partial{xq_{s}(x,Q^{2})}}{\partial{\ln}Q^{2}}=P_{qg}{\otimes}xg+P_{qq}{\otimes}xq_{s}
-\frac{27\alpha_{s}^{2}(Q^{2})}{160\mathcal{R}^{2}Q^{2}}[xg(x,Q^{2})]^{2}+\mathrm{G_{HT}},
\end{eqnarray}
where the higher dimensional gluon term $\mathrm{G_{HT}}$ is here
assumed
to be zero.\\
The standard DGLAP evolution equation for singlet and gluon
distributions has the following forms:
\begin{eqnarray}
\frac{\partial{xg(x,Q^2)}}{\partial{\ln{Q^2}}}|_{\mathrm{DGLAP}}=P_{gq}{\otimes}xq_{s}+P_{gg}{\otimes}xg=
\frac{\alpha_{s}(Q^2)}{2\pi}
\int_{x}^{1}\frac{dz}{z^2}x\bigg{[}P_{gq}(\frac{x}{z})xq_{s}(z,Q^2)+P_{gg}(\frac{x}{z})xg(z,Q^2)\bigg{]},
\end{eqnarray}
\begin{eqnarray}
\frac{\partial{xq_{s}(x,Q^2)}}{\partial{\ln{Q^2}}}|_{\mathrm{DGLAP}}=P_{qq}{\otimes}xq_{s}+P_{qg}{\otimes}xg=
\frac{\alpha_{s}(Q^2)}{2\pi}
\int_{x}^{1}\frac{dz}{z^2}x\bigg{[}P_{qq}(\frac{x}{z})xq_{s}(z,Q^2)+P_{qg}(\frac{x}{z})xg(z,Q^2)\bigg{]},
\end{eqnarray}
where $P_{ij}^{,}$s are the splitting functions in the desired
order in $\alpha_{s}$.\\


\section{References}
1. M.Klein, Annalen Phys.{\bf528}, 138 (2016).\\
2. K.J.Eskola et al., Nucl.Phys.A {\bf661}, 645 (1999).\\
3. K.J.Eskola et al., arXiv:0110348 (2001).\\
4. H.Paukkunen, K.J.Eskola and N.Armesto, arXiv:1306.2486 (2013),
XXI International Workshop on Deep-Inelastic Scattering and
Related Subjects.\\
5. P.Agostini  et al. [LHeC Collaboration and FCC-he Study Group ], J. Phys. G: Nucl. Part. Phys. {\bf48}, 110501 (2021).\\
6. B.Kopeliovich, J.Raufeisen and A.Tarasov, Phys.Rev.C {\bf62},
035204 (2000).\\
7. J.Raufeisen, arXiv:0204018 (2002).\\
8. B.Kopeliovich and B.Povh, arXiv:9504380 (1995).\\
9. M.Arneodo et al. [New Muon Collaboration], Nucl.Phys.B
{\bf481}, 23 (1996).\\
10. E.R.Cazaroto et al., Phys.Lett.B {\bf669}, 331 (2008).\\
11. P.Paakkinen, arXiv:1802.05927.\\
12. I.Helenius, M.Walt and W.Vogelsang, Phys.Rev.D {\bf105}, 094031 (2022).\\
13. J.Ethier and E.R.Nocera, Annu. Rev. Nucl. Part. Sci. {\bf70},
43 (2020).\\
14. M.Hirai, S.Kumano and M.Miyama, Phys. Rev. D {\bf64}, 034003 (2001).\\
15. L.N. Lipatov, Sov. J. Nucl. Phys.{\bf20}, 94 (1975).\\
16. V.N. Gribov, L.N. Lipatov, Sov. J. Nucl. Phys.{\bf15}, 438
(1972).\\
17. G. Altarelli, G. Parisi, Nucl. Phys. B{\bf126}, 298 (1977).\\
18. Yu.L. Dokshitzer, Sov. Phys. JETP {\bf46}, 641 (1977).\\
19. J.Sheibani, A.Mirjalili and S.Atashbar Tehrani,
Phys. Rev. C {\bf98}, 045211 (2018).\\
20. I. Sick and D. Day, Phys. Lett. B {\bf274}, 16 (1992).\\
21. A.Accardi et al., Eur.Phys.J.A {\bf52}, 268 (2016).\\
22. R. Abdul Khalek et al., (2021), arXiv:2103.05419
[physics.ins-det].\\
23. L. V. Gribov, E. M. Levin and M. G. Ryskin, Phys. Rept.
{\bf100}, 1 (1983).\\
24. A. H. Mueller and J. w. Qiu, Nucl. Phys. B {\bf268}, 427
(1986).\\
25. W. Zhu and J. h. Ruan, Nucl. Phys. B {\bf559}, 378 (1999).\\
26. A.M.Stasto, Acta Phys.Polon.B {\bf33}, 1571(2002).\\
27. S.Zarrin and S.Dadfar, Phys.Rev.D {\bf105}, 094037 (2022).\\
28.  K. Prytz, Eur. Phys. J. C {\bf22}, 317 (2001).\\
29.  M. Lalung, P. Phukan and J. K. Sarma, Nucl. Phys. A {\bf984},
29 (2019).\\
30. M. Devee and J. K. Sarma, Nucl. Phys. B {\bf885}, 571
(2014).\\
31. G.R.Boroun, Eur.Phys.J.C {\bf81}, 851 (2021).\\
32. G.R.Boroun, Eur.Phys.J.Plus {\bf137}, 259 (2022).\\
33. J.Rausch, V.Guzey and M.Klasen, Phys.Rev.D {\bf107}, 054003
(2023).\\
34. H.Khanpour and S.Atashbar Tehrani, Phys. Rev. D {\bf93}, 014026 (2016).\\
35. H.Khanpour et al., Phys.Rev.D {\bf 104}, 034010 (2021).\\
36. S. Atashbar Tehrani, Phys. Rev. C {\bf86}, 064301 (2012).\\
37. M. Hirai, S. Kumano and T.-H. Nagai, Phys. Rev. C {\bf70},
044905 (2004).\\
38. S. Atashbar Tehrani, A. N. Khorramian and A. Mirjalili, Int.
J. Mod. Phys. A {\bf20}, 1927 (2005).\\
39. J.Kwiecinski et al., Phys.Rev.D {\bf42}, 3645 (1990).\\
40. L.Frankfurt et al., Rep. Prog. Phys. {\bf85}, 126301 (2022).\\
41. G.R.Boroun, M.Kuroda and D.Schildknecht, arXiv[hep-ph]:2206.05672.\\
42. G.R.Boroun, arXiv[hep-ph]: 2312.04228.\\
43. T.J.Hou et al., Phys.Rev.D {\bf103}, 014013 (2021).\\
44. H. Khanpour, A. N. Khorramian and S. Atashbar Tehrani, J.Phys.G {\bf40}, 045002 (2013).\\
45. P. Jimenez-Delgado and E. Reya, Phys.Rev.D {\bf79} , 074023
(2009).\\
46. D.I.Kazakov and A.V.Kotikov, Phys.Lett.B{\bf291}, 171(1992).\\
47. E.B.Zijlstra and W.L.van Neerven, Nucl.Phys.B{383}, 525(1992).\\
48. R. Brock et al. [CTEQ], Rev. Mod. Phys. {\bf67}, 157
(1995).\\
49. S. Moch, J.A.M. Vermaseren, and A. Vogt, Phys. Lett. B
{\bf606}, 123 (2005).\\
50. M.Gl$\mathrm{\ddot{u}}$k, C.Pisano and E.Reya,
Phys.Rev.D{\bf77}, 074002 (2008).\\
51. A. Vogt, S. Moch and J.A.M. Vermaseren, Nucl.Phys.B {\bf691},
129 (2004).\\
52. B.G.Shaikhatdenov, A.V.Kotikov, V.G.Krivokhizhin and
G.Parente, Phys.Rev.D {\bf81}, 034008 (2010).\\
53. G.R.Boroun and B.Rezaei, Phys.Rev.D {\bf105}, 034002 (2022).\\
54. M. M. Block, L. Durand and P. Ha, Phys. Rev. D {\bf89}, 094027
(2014).\\
55. M. Froissart, Phys. Rev. {\bf123}, 1053 (1961).\\
56. F.D. Aaron et al., [H1 and ZEUS Collaborations], JHEP
{\bf1001}, 109 (2010).\\
57. L.P.Kaptari et al., Phys.Rev.D{\bf 99}, 096019 (2019).\\
58. M.Botje, QCDNUM16: A fast QCD evolution program, 1997,
https://www.nikhef.nl/h24/qcdcode/qcdnum1612.pdf.\\
59. K. Kovarik et al., Phys.Rev.D {\bf93}, 085037 (2016).\\
60. D.P.Anderle, et al., Front. Phys. {\bf16},  64701 (2021).\\




\end{document}